\documentclass[prd,nofootinbib,amsfonts,notitlepage,showpacs]{revtex4-1}

\usepackage{dcolumn}
\usepackage{amssymb}
\usepackage{graphicx,bm,amsmath,color}
\usepackage{hyperref}
\hyperbaseurl{http://arxiv.org/abs} 

\usepackage{bm}
\usepackage{enumerate}

\def\be{\begin{enumerate}}                     \def\ee{\end{enumerate}}
\def\beq{\begin{equation}}                     \def\eeq{\end{equation}}

\newcommand{\bea}{\begin{eqnarray}}   
\newcommand{\eea}{\end{eqnarray}}

\def\3halfs{\textstyle{\frac{3}{2}}}                       

\def\ben{\begin{enumerate}}                   \def\een{\end{enumerate}}
\def\bitem{\begin{itemize}} \def\eitem{\end{itemize}}

\newcommand{\comments}[1]{}

\begin{document}
\title{On the use of energy loss mechanisms to constrain Lorentz invariance violations\\ 
}
\author{Diego Maz\'on} 
\email{dg.mazon@gmail.com}
\affiliation{2150 W 39\textsuperscript{th} Avenue \\Vancouver, BC, V6M1T5\\ Canada}

\begin{abstract}

In light of recent and probably upcoming observations of very high energy astroparticles, such as those reported by the \textit{IceCube} collaboration,  we readdress the energy loss mechanism by Lorentz violating particles. We analytically show that \textit{Cohen-Glashow's formula}  for  energy loss is connected with a Poisson distribution for the number of decays, whose large fluctuations prevent placing bounds on Lorentz invariance violations. However, this model ignores the sharp change in the decay width after each process. We propose replacing Poisson statistics with a new distribution that takes this into account. We study the average final energy and its fluctuations according to the new statistics, contrasting it with Cohen-Glashow's result, and discussing the reliability of energy loss mechanisms to constrain violations of Lorentz invariance.


\end{abstract}
\maketitle




%


\section{Introduction}

Placing limits on Lorentz invariance violating models has a twofold importance. On the one hand, Lorentz invariance is a cornerstone of the standard model of elementary interactions that enormously restricts the  set of possible theories beyond the standard model -- especially when imposed along with unitarity. Thus, constraining deviations from this symmetry is a key issue in the development of new physical theories. Furthermore, because of the infinite volume of the Lorentz group, this is suppose to be an endless research activity, unless a departure from this symmetry is ever found. On the other hand, from a more specific perspective, deviations from exact Lorentz invariance are motivated by some theoretical indications in several approaches to a quantum theory of gravity, ranging from string theory (see Ref.~\cite{Kostelecky:1988} for the first article and Ref.~\cite{Mavromatos:2007} for a review) to condensed matter analog models of gravity (see Ref.~\cite{Barcelo:2005}), just to mention a few of them (see Ref.~\cite{Liberati:2013,AmelinoCamelia:2008,Mattingly:2005} for reviews and extensive lists of references). An heuristic argument for the existence of such departures from Lorentz invariance is the presence of a fundamental length scale, related to the Planck length, in the physical regime where both quantum mechanics and gravitation must be taken into account, such as possible generalizations of the  Heisenberg uncertainty principle (see Refs.~\cite{Garay:1994, Hossenfelder:2012}).

The insurmountable gap between the Planck energy and the accessible energies in man-made accelerators has made phenomenologists look for situations in which an amplifying mechanism makes a minute effect detectable with the help of a high precision measurement~\cite{Ellis:1983,Mercati:2010,Berchera:2013} or use the universe as a natural accelerator~\cite{GonzalezMestres:1997,Coleman:1998,Aloisio:2000,Bietenholz:2008}. As a recent example, the IceCube collaboration has reported the observation of 28 neutrinos with energies ranging from teens to one thousand of TeV~\cite{IceCube:Science,IceCube:PRL}. This observation has been used to put stringent bounds on Lorentz invariance violations in Refs.~\cite{Borriello:2013,Stecker:2013}.

Deviations from exact Lorentz invariance may be classified according to two different properties: i)~Whether they preserve a relativity principle or not. The formers are said to be deformations of Lorentz invariance, whereas in the latter case, the departures are classified as violations. ii)~Whether the energy-momentum conservation law remains additive or becomes non-linear. In this work, we will readdress one of the most efficient means to constrain Lorentz invariance violations  with a standard, additive energy-momentum conservation law: energy loss mechanisms. 

Decays that are forbidden in Lorentz invariant theories may become permitted if this symmetry is not exact. Then, if a particle decays into itself plus other particles, which we will hereafter refer to as  `radiated particles', the symmetry breaking makes particles lose energy.  If the energy loss increases with the amount of Lorentz invariance violation, the observations of high energy particles may be used to place limits on the size of the Lorentz invariant violating parameters (see Refs.~\cite{Cowsik:2012,Diaz:2013,Stecker:2013,Borriello:2013,Cohen:2011,Carmona:2012,
Borriello:2013,Maccione:2011}). Reactions of this sort include: photon and neutrino splitting ($\gamma\rightarrow 3\gamma$ and $\nu_l\rightarrow \nu_l\, \nu_{l'}\, \bar\nu_{l'}$), photon production by charged leptons and neutrinos ($e^{\pm}\rightarrow e^{\pm}\, \gamma$ and $\nu_l\rightarrow \nu_l\, \gamma$), and pair production by neutrinos ($\nu_l\rightarrow \nu_l\, e^{+}\, e^{-} $), which will be the one we will use hereafter when specific information is needed; we refer the reader to Ref.~\cite{Carmona:2012} for all details about this decay. This latter process is only permitted if the incoming neutrino is superluminal\footnote{We use the word  `superluminal' in a somewhat vague sense. What we strictly mean is that the maximum speed of propagation of neutrinos must be larger than the maximum speed of the radiated particles (electrons and positrons), for which we will assume a special-relativistic dispersion relation. Whether or not neutrinos are faster than light is not relevant for this process.} and its energy is above certain threshold. The presence of such finite threshold energy  is a simple indication of the violation of Lorentz invariance, as energy transforms under Lorentz transformations.

The relation among  energy loss,  decay width, and  path length so far used to constrain Lorentz invariance violations was given for the first time in Ref.~\cite{Cohen:2011}.  After computing the energy loss in each decay in the following section, we will devote Sec.~\ref{sec:CGP} to show that  Cohen-Glashow's result is connected with the assumption that the number of decays along the flight path follows a Poisson distribution. We will calculate the uncertainty in the energy due to the fluctuations in the number of decays according to Poisson statistics, showing that they are so large that they prevent  placing bounds on Lorentz invariance breaking from the observation of high-energy particles coming from far enough sources. In Sec.~\ref{sec:realistic}, we will provide a more realistic statistics for the number of events that gives a different average energy and  much more restricted fluctuations, which do allow the use of energy loss mechanisms to constrain Lorentz invariance. The fluctuations in energy loss in each decay will be studied in Sec.~\ref{sec:each_decay}, whereas Sec.~\ref{sec:generality} will be devoted to consider the generality of the results. We will finish with a summary and a discussion.

\section{Mean energy loss in each decay}
Throughout this paper, we will deal with two independent sources of fluctuations. The first one is due to the quantum probabilistic nature of the decays: the number of decays in a given period of time or a flight path is not a fixed number, but a random one that follows a probabilistic distribution. We will write the symbol $\langle \quad \rangle$ to denote  average values with respect to this distribution. The second one has its origin on the fact that the daughter particles may distribute the parent particle's energy in several ways; therefore, the energy loss in each decay is not always the same, but it follows a probability distribution. Mean values related to this distribution, such as the one in the following equation, will be denoted by an overline $\bar{\quad}\,$. 

The average energy radiated by a particle with incoming energy $E$ in each process is obtained by averaging with the differential decay width $\Gamma(E,E')$ 
\begin{equation}
\begin{aligned}
\label{energyradiated}
\overline{\delta  E} &= \dfrac{\int (E'-E)\; \Gamma (E,E')\; dE' }{\int \Gamma (E,E')\; dE'}   \\
 &=\dfrac{\int (E-E')\; \Gamma (E,E')\; dE' }{\Gamma (E)}=-k\, E \,,
 \end{aligned}
\end{equation}
where $\Gamma (E)$ is the decay width and acts as a normalization of the distribution $\Gamma(E,E')$, and $k$ is a dimensionless pure number between 0 and 1, which depends on the specific dynamical matrix element and the dispersion relation for free particles; see Fig.~\ref{fig:k_m} for the specific case of pair production by superluminal neutrinos. The last equality is valid for a wide range of differential decay widths, as we will show in Sec.~\ref{sec:each_decay}. From now on, we will assume that the energy loss in each decay is a non-fluctuating number that equals the average value given in Eq.~(\ref{energyradiated}). In Sec.~\ref{sec:each_decay} we will return to this issue by studying the fluctuations around the mean value. 

\section{The Cohen-Glashow model matches Poisson statistics}
\label{sec:CGP}
The mean neutrino energy after $n$ decays will be $(1-k)^n$ times the initial energy. After having travelled a given distance $x$, the number of decays will be a fluctuating number. If we consider that the decay of a neutrino is independent of the number of previous decays, then we can obtain the final mean energy averaging the energy after $n$ decays, $\bar{E}_n\,$, with a Poisson probability distribution
\begin{equation}
\begin{aligned} 
\label{E(N)}
\langle \bar{E} (x) \rangle &= E_i\, \sum_{n=0}^{\infty}\, (1-k)^n\,\dfrac{e^{-N}\, N^n}{n!}  \\
&= E_i\, e^{-k\,N}\, ,
\end{aligned}
\end{equation}
where $E_i\equiv E(x=0)$ is the initial energy and $N$ is the mean number of decays within this approximation and can be related to the distance of flight $x$  
\beq
\begin{aligned}
x&=\int_0^x dx' =\int _0^{N} \frac{dN'}{\Gamma (\langle E \rangle)} \\
&=\dfrac{1}{\Gamma (E_i)}\, \int _0^{N} e^{(5+3m)\,k\,N'} \, dN'\,,
\end{aligned}
\label{x}
\eeq
where $5+3\,m$ is the power of the energy in the decay rate 
\begin{equation}
\Gamma (E)\propto E^{5+3m}\, ,
\label{prop}
\end{equation}
and we have used this particular notation because it matches the decay rate for $\nu_l\rightarrow \nu_l\, e^{+}\, e^{-}$ when the dispersion relation for free neutrinos is $E=p\, (1+\eta_m\,(p/\Lambda)^m)$; see Ref.~\cite{Carmona:2012}. 

Performing the integral in Eq.~(\ref{x}) and solving for $N$, we obtain
\begin{equation}
\label{N(x)}
N= \frac{\ln \left[1+(5+3m)\, k\,\Gamma (E_i)\, x\, \right]}{(5+3m)\,k}\,.
\end{equation}
Replacing $N$ in Eq.~(\ref{E(N)}), we recover the expression for the final neutrino energy given in Ref.~\cite{Cohen:2011} (for the particular case $m=0$)
\beq
\begin{aligned}
\label{finalenergy}
\langle \bar{E} (x) \rangle & = E_i\, \left[1+(5+3m)\, k\;\Gamma (E_i)\, x \, \right]^{-1/(5+3m)}  \\
& =E_i\, \left[1+\left(E_i/E_t\right)^{5+3m} \right]^{-1/(5+3m)}\, ,
\end{aligned} 
\eeq
where $E_t$ is the so-called terminal energy
\begin{equation}
E_t^{5+3m}\equiv\frac{E_i^{5+3m}}{(5+3m)\,k\, \Gamma (E_i)\, x}\; ,
\label{E_t}
\end{equation}
or, in words, the terminal energy is the value of the final energy when this is much lower than the initial energy. Note that according to Eq.~(\ref{prop}) the terminal energy does not depend on the initial energy; thus, we may replace the initial energy in the right-hand-side of Eq.~(\ref{E_t}) with any other value of the energy. Given the equivalence between the Cohen-Glashow result for the energy loss and the one computed in this section using a Poisson distribution,  we will refer to this model as the Cohen-Glashow-Poisson (CGP) model from now on.

Mean values are, however, not enough; the statistical fluctuations, or at least their estimates, are required to place sensible constraints on Lorentz invariance violations based on a probabilistic phenomenon like particle decay.  If within this model we compute the standard deviation in the final  energy owing to the fluctuation in the number of decays, we obtain
\begin{equation}
\begin{aligned}
\label{sdpoisson}
\sigma ^2 _{E}(x) &= \langle  \bar{E}^2 (x) \rangle - \langle \bar{E} (x) \rangle ^2   \\ 
&=E_i^2\,e^{(1-k)^2\,N-N} - \langle \bar{E} (x) \rangle ^2  \\ 
&=\langle \bar{E} (x) \rangle ^2\, \left(e^{k^2\,N}-1\right)  \\ 
&=  \langle \bar{E} (x) \rangle ^2\, \left[\left(\dfrac{E_i}{\langle \bar{E} (x) \rangle}\right)^{k}-1\right]  \\
&\simeq  \langle \bar{E} (x) \rangle ^2\, \left[\left(\dfrac{E_i}{E_t}\right)^{k}-1\right]  \, ,
\end{aligned}
\end{equation}
where in the fourth line we have made use of Eq.~(\ref{E(N)}) and in the last step we have used Eq.~(\ref{finalenergy}) and have assumed that the initial energy is at least a few times larger than the terminal energy. 
This result would call into question recent analyses that only make use of the final average energy, but not of the fluctuations.
And particularly those that are based on the observation of only a few events of high energy particles or those in which the initial energy is unknown, as the fluctuations grow with the initial energy. In Ref.~\cite{Maccione:2011}, on the contrary, the data are Monte Carlo simulated, and it is pointed out that the final energy in Eq.~(\ref{finalenergy}) does not have a sharp cut-off in the spectrum because of the finiteness of the baseline experiment (note, however, that according to Eqs.~(\ref{finalenergy})~and~(\ref{sdpoisson}), the ratio between the standard deviation and the mean increases as the path length grows). 

\begin{figure}
\centerline{\includegraphics[width=0.47\textwidth]{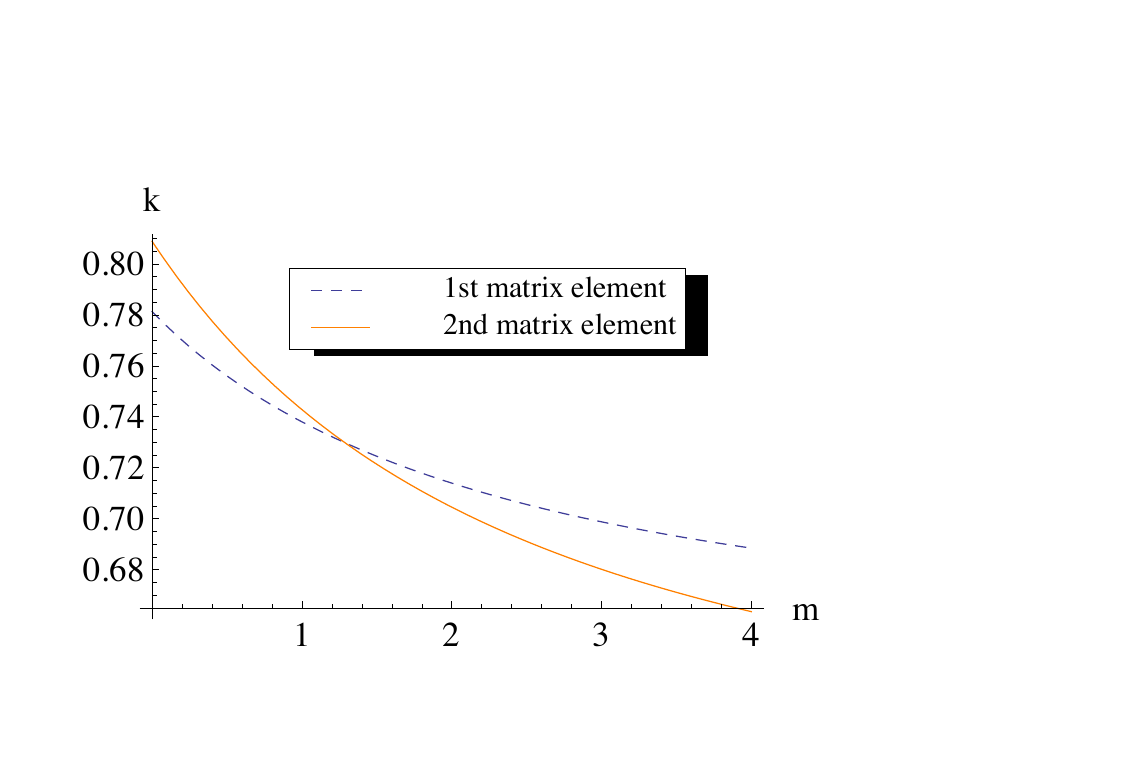}}
\caption{Values of $k$ for the process $\nu_l\rightarrow \nu_l\, e^{+}\, e^{-} $, for two different matrix elements and different superluminal dispersion relations $E=p\, (1+\eta_m\,(p/\Lambda)^m)$  (see Ref.~\cite{Carmona:2012}, where $k$ is $\xi'/\xi$ for the first matrix element and $\tilde\xi'/\tilde\xi$ for the second one).}
\label{fig:k_m}
\end{figure}

Nonetheless, the model based on Poisson statistics  studied in this section does not consider the fact that the variation of the decay rate  with energy is entangled with the number of previous decays. Intuitively, this approximation, in which the dependence of the decay rate on energy is separated from the  energy loss in each decay, should work only for those cases where in each disintegration a little amount of energy is radiated and the rate of decays $dN/dx$ is high for a given initial energy; hence, according  to Eqs.~(\ref{energyradiated})~and~(\ref{N(x)}), the approximation is valid as long as $k\ll1$. In this limiting case, the mean value of the Poisson distribution is $N\simeq \Gamma (E_i)\, x$. However, this is not necessarily the case, and it is definitely not for the production of leptonic pairs by superluminal neutrinos. In Fig.~\ref{fig:k_m} we can see that the values of $k$ range from $0.7$ to $0.8\,$. 

We can see the inaccuracy of the Cohen-Glashow-Poisson model from another angle. In Refs.~\cite{Cohen:2011,Carmona:2012,Maccione:2011}, the left-hand side of Eq.~(\ref{energyradiated}) over the mean time  between decays,
\begin{equation}
\overline{\delta E}/\tau (E)\equiv\overline{\delta E} \cdot \Gamma (E)\, ,
\end{equation}  
is identified with the instantaneous rate of change of energy, $dE/dt$. After integrating the differential equation,  Eq.~(\ref{finalenergy}) is obtained. However, this identification is only correct if $\tau (E)$ is much lower than the flight path~$x$ for all the energies the particle may have along the path. As the minimum energy is the final energy, the relevant condition turns to be $\tau (E (x))\ll x$. Note that as the flight path grows, the final energy decreases, and $\tau (E (x))$ increases; therefore, the precedent condition is not equivalent to only consider  long enough distances. More precisely,
\begin{equation}
\Gamma (E(x))\,x=\frac{\Gamma (E_i)\, x}{1+(5+3\,m)\,k\,\Gamma (E_i)\,x}\,,
\end{equation}
so that $\tau (E (x))\ll x$ (i.e., the validity of Cohen-Glashow-Poisson approximation) requires that the radiated energy in each emission is very low, $k\ll 1\,$, besides distances much longer than the initial free path, $\Gamma (E_i)\, x \gg 1$. 

\section{Distribution of decays for an energy-dependent decay rate}
\label{sec:realistic}

In this section, we will take into consideration that after each emission the particle loses part of its energy, and this instantaneously affects its decay rate, while during the travel between emissions, the energy and the decay rate remain constant. The probability distribution we propose is given by
\beq
\begin{aligned} 
\mathcal{P} (n; x) & = \int_0^x \mathcal{P} (n-1; s)\, \Gamma (\bar{E}_{n-1})\, ds \, \mathcal{Q} (n; x-s) \, ,
\end{aligned}
\label{realistic1}
\eeq
with $n=1, 2, 3, ...$ and $\bar{E}_n\equiv E_i\, (1-k)^n$ the energy after $n$ decays. That is, in words, the probability of $n$ decays between the points $0$ and $x$ along the flight path ($\mathcal{P} (n; x)$) is given by the probability of $n-1$ decays between $0$ and $s$, times the likelihood of an $n^{\text{\tiny th}}$ decay between $s$ and $s+ds$ ($\Gamma (\bar{E}_{n-1})\, ds $), times the probability of no decay between $s$ and $x$ when $n$ decays have taken place before ($\mathcal{Q} (n; x-s)$), integrated to all possible values of $s$ from $0$ to $x$. Taking into consideration that the probability of no decay between $0$ and $x+dx$  after $n$ decays is given by the probability of no disintegration between $0$ and $x$ times the probability of no decay between $x$ and $x+dx$, which is, in turn, $1$ minus the probability of a decay between $x$ and $x+dx$, we obtain 
\begin{equation}
\mathcal{Q} (n; x+dx)=\mathcal{Q} (n; x)\, \left(1-\Gamma (\bar{E}_{n})\, dx\right)\, ;
\end{equation}
thus,
\begin{equation}
\frac{d\,\mathcal{Q} ( n; x)}{dx}=-\,\Gamma (\bar{E}_{n})\,\mathcal{Q} (n; x)\, .
\end{equation}
Taking into account that there is no decay at the initial point $\mathcal{Q} (n; x= 0)=1$, we obtain
\begin{equation}
\mathcal{Q} (n; x)=e^{-\Gamma (\bar{E}_{n})\, x}\, .
\end{equation}
And we can then rewrite  Eq.~(\ref{realistic1}) as
\begin{align} 
\mathcal{P} (n; x) = \Gamma (\bar{E}_{n-1}) \, e^{-\Gamma (\bar{E}_{n})\,x} \int_0^x \mathcal{P} (n-1; s)\, e^{\Gamma (\bar{E}_{n})\,s}\, ds\, .
\label{realistic}
\end{align}
This recurrence relation together with the initial condition
\begin{equation}
\mathcal{P} (n=0; x)\equiv \mathcal{Q} (n=0; x)=e^{-\Gamma (E_i)\, x}\, ,
\label{initial_condition}
\end{equation}  
define the desired normalized probability distribution that will replace Poisson's in this work and are the central equations of this paper.\footnote{Although this recurrence relation only involves trivial integrals of exponentials, the closed-form solution (the general term  as an explicit function of $n$) is given by a cumbersome integral.}  In the limiting case $k\rightarrow 0$, the distribution $\mathcal{P} (n; x)$ tends to a Poisson distribution with mean value $\Gamma (E_i)\,x$; therefore, the Cohen-Glashow-Poisson model and the one presented in this section converge in this limit, in which the energy of the radiated particles tends to zero. 
On the other hand, let us remark that for an arbitrary $k$, the probability of each independent decay is different from the one given by the Poisson distribution corresponding to the instantaneous neutrino decay rate (note that this Poisson distribution is not the one in Sec.~\ref{sec:CGP} that matches the Cohen-Glashow model). That is, if one knows with certainty that $n-1$ decays have taken place at point $x_1$, the probability that a single $n^{\text{\tiny th}}$ decay takes place between $x_1$ and $x_2$ is given by 
\begin{equation}
\begin{aligned} 
\label{remark}
&\int_{x_{1}}^{x_{2}} \mathcal{Q} (n-1; s-x_{1}) \, \, \Gamma (\bar{E}_{n-1})\, ds \, \mathcal{Q} (n; x_{2}-s)  \\ 
&=\Gamma (\bar E_{n-1})\, \dfrac{e^{-\Gamma (\bar E_{n})\delta x}-e^{-\Gamma (\bar E_{n-1})\delta x}}{\Gamma (\bar E_{n-1})-\Gamma (\bar E_{n})}\, ,
\end{aligned}
\end{equation}
with $\delta x\equiv x_2-x_1$, whereas according to the Poisson distribution corresponding to the instantaneous decay rate, the probability is
\begin{equation}
\Gamma (\bar E_{n-1})\,\delta x\, e^{-\Gamma (\bar E_{n-1})\delta x}\, ;
\end{equation}
therefore, they only agree (for a general $k$) when $\delta x$ is infinitesimal, where the agreement is expected as the probability is just the definition of decay rate. As the probability of no decay in a finite interval directly follows from the definition of decay rate, both distributions agree on the probability of zero decays, but not on the probability of a specific (non-zero) number of decays in a finite interval.  

According to the distribution of decays for an energy-dependent decay rate, the average final energy is
\begin{equation}
\langle \bar{E} (x) \rangle = E_i \sum_{n=0}^{\infty} (1-k)^n\, \mathcal{P} (n; x) \, ,
\end{equation}
and the uncertainty due to the fluctuations in the number of decays may be characterized by the standard deviation
\begin{equation}
\begin{aligned}
\sigma^2_{E(x)} & =\langle \bar{E}^2 (x) \rangle - \langle \bar{E} (x) \rangle ^2 \\
& = E_i^2 \sum_{n=0}^{\infty} (1-k)^{2n}\, \mathcal{P} (n; x) - \langle \bar{E} (x) \rangle ^2 \, .
\end{aligned}
\end{equation}

\begin{figure}
\centerline{\includegraphics[width=0.47\textwidth]{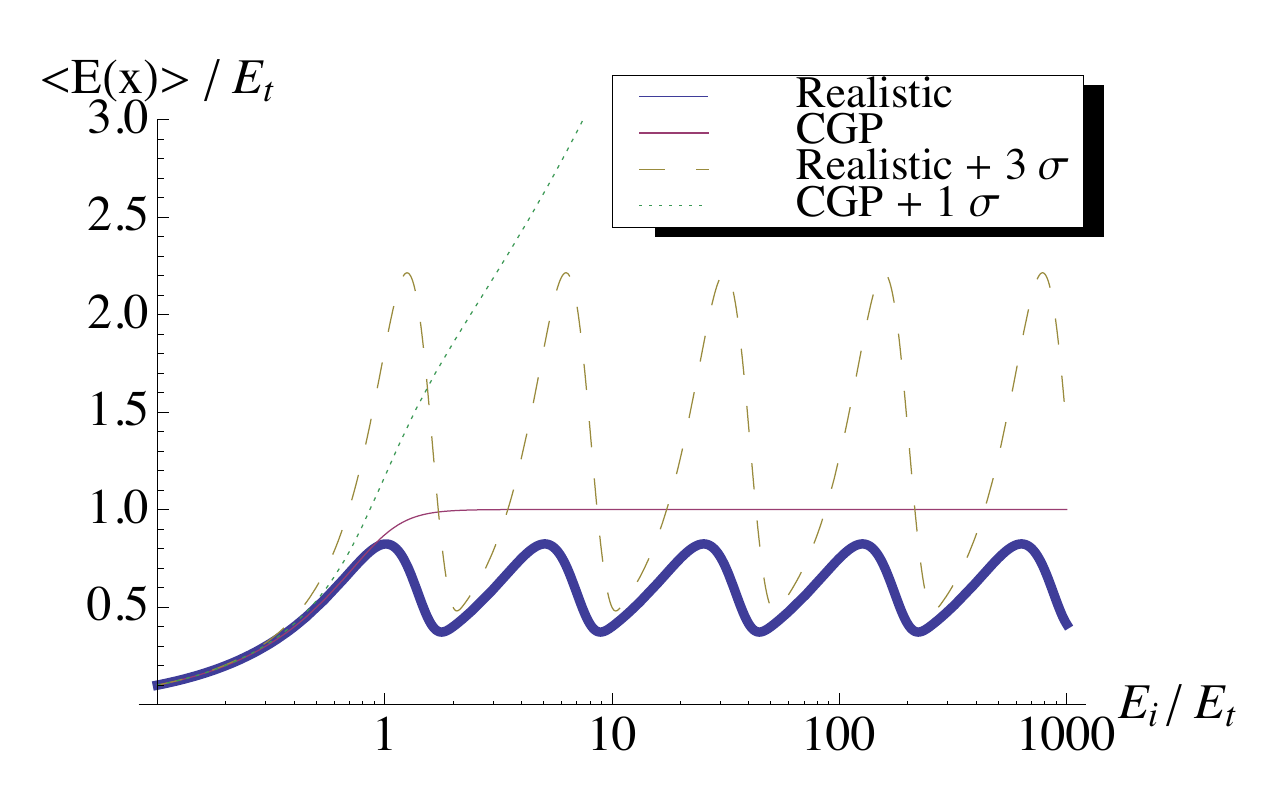}}
\caption{Final energy after a given flight path (in terms of the terminal energy for this distance) against initial energy (in terms of terminal energy as well) for an energy independent velocity  ($E=p\, (1+\eta_0)$) and $k=0.8$ (so that it approximately applies to the two dynamical matrix elements considered in Fig.~\ref{fig:k_m}). Note that the horizontal axis uses logarithmic scaling.}
\label{fig:Efinal_Einitial}
\end{figure}

\begin{figure}
\centerline{\includegraphics[width=0.47\textwidth]{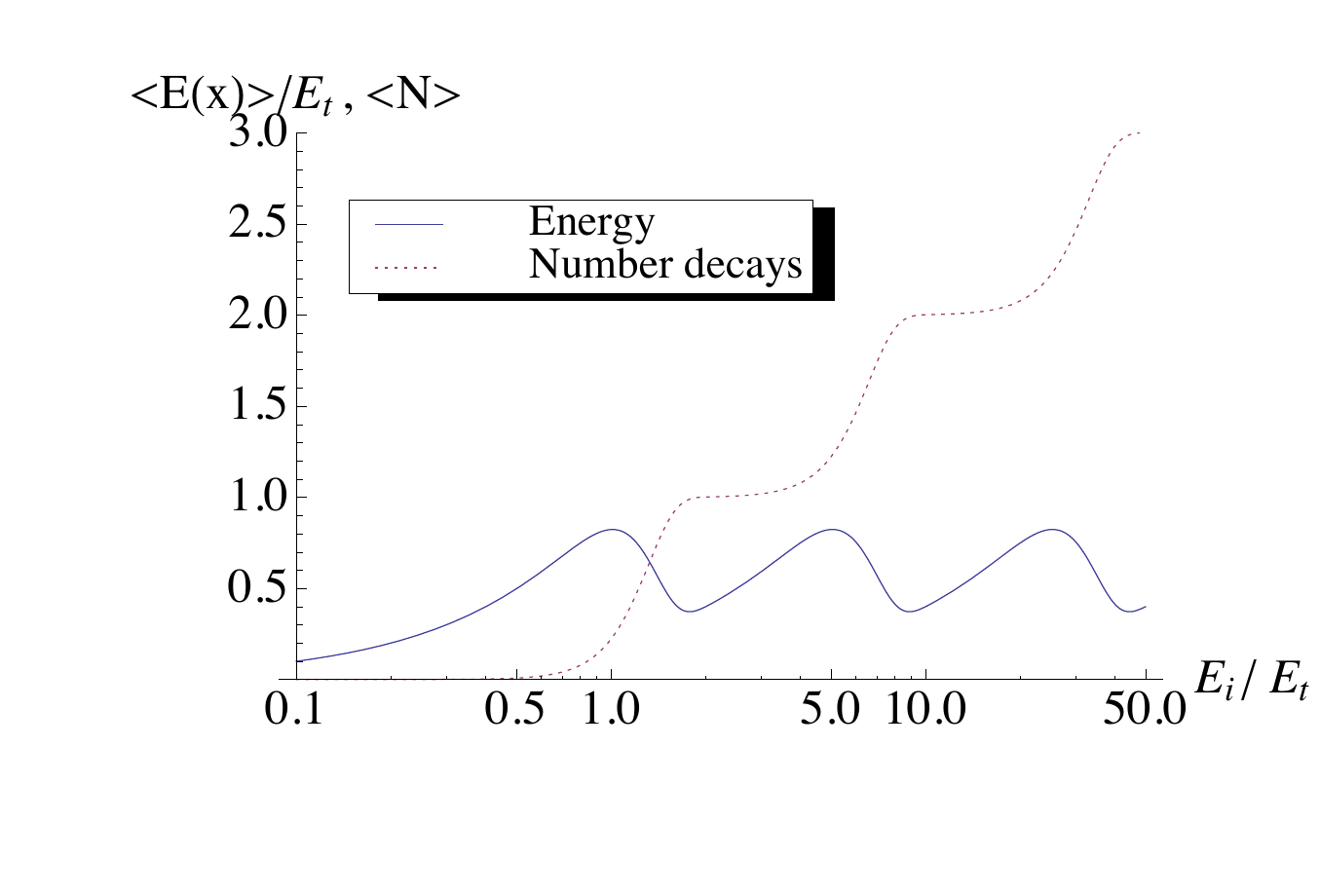}}
\caption{Average final energy after a given distance travelled (in terms of the terminal energy for this given distance) and average number of decays against initial energy (in terms of terminal energy as well) for an energy independent velocity  ($E=p\, (1+\eta_0)$) and $k=0.8$ (so that it approximately applies to the two dynamical matrix elements considered in Fig.~\ref{fig:k_m}). Note that the horizontal axis uses logarithmic scaling.}
\label{fig:NumberDecays}
\end{figure}

In Fig.~\ref{fig:Efinal_Einitial} the final energy within this distribution as a function of the initial energy is contrasted with the Cohen-Glashow-Poisson model for the pair production by superluminal neutrinos process $\nu_l\rightarrow \nu_l\, e^{+}\, e^{-} $.  We see that both average energies agree each other for initial energies lower than the terminal energy. However, for larger initial energies, whereas in the Cohen-Glashow-Poisson model the mean final energy (thin magenta line) reaches a constant value (the terminal energy), in the more realistic model presented in this work, the mean final energy (thick blue line) oscillates between approximately $0.4$ and $0.8$ times the terminal energy. The reason why the final energy decreases in some intervals is that the increase in the initial energy is not able to balance out the energy loss due to a new decay (see Fig.~\ref{fig:NumberDecays}).  Regarding fluctuations, the standard deviation monotonically grows with the initial energy (see Eq.~(\ref{sdpoisson})) in the Cohen-Glashow-Poisson model (dotted line), while it is bounded in the realistic model (dashed line). Specifically, the average final energy plus three standard deviations is lower than 2.5 times the terminal energy.

\begin{figure}
\centerline{\includegraphics[width=0.48\textwidth]{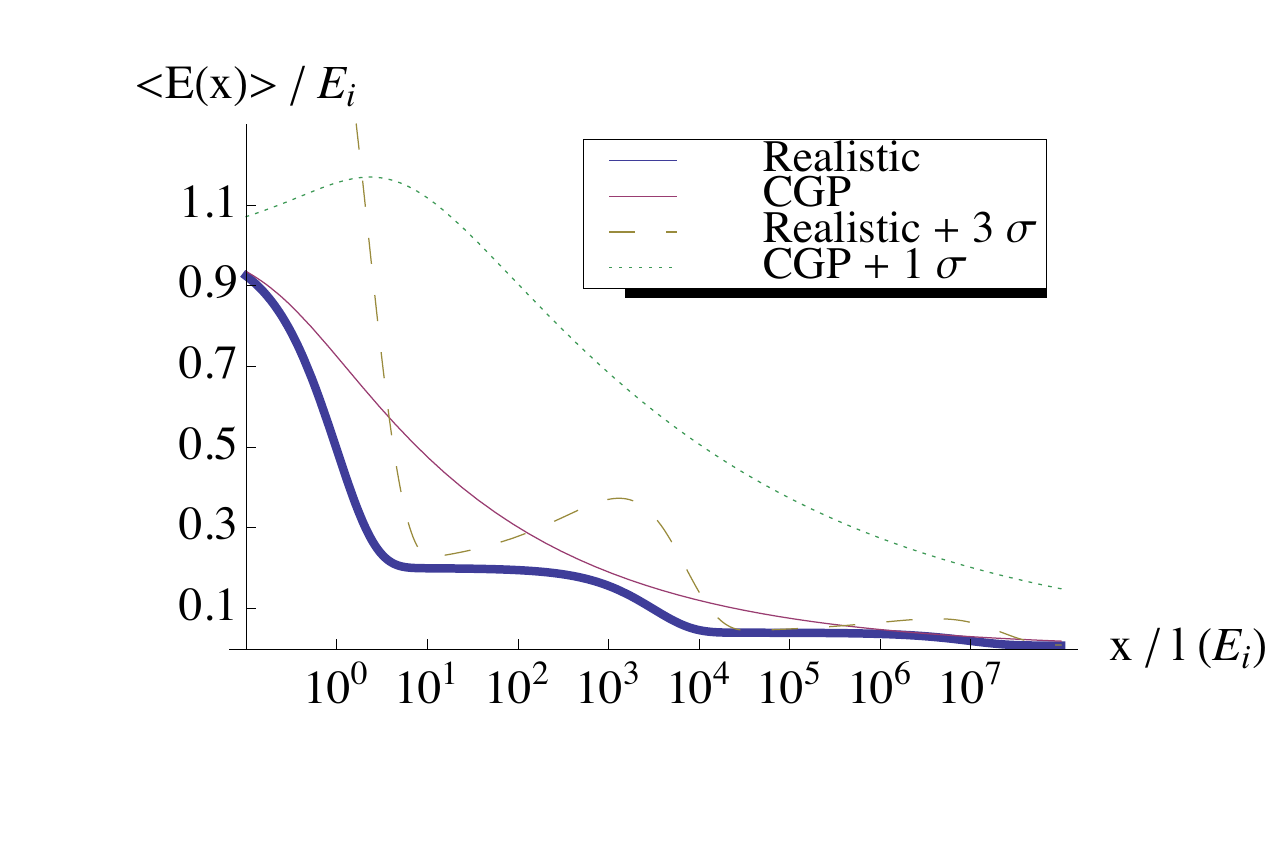}}
\caption{Energy  (in terms of a given initial energy) against distance travelled $x$ (in terms of the initial mean free path $l (E_i)\equiv \Gamma ^{-1} (E_i)$) for an energy-independent speed  ($E=p\, (1+\eta_0)$) and $k=0.8$ (so that it approximately applies to the two dynamical matrix elements considered in Fig.~\ref{fig:k_m}). Note that the horizontal axis uses logarithmic scaling.}
\label{fig:Efinal_distance}
\end{figure}

Alternatively, in Fig.~\ref{fig:Efinal_distance} the energy is plotted against the flight path travelled. We can notice that the Cohen-Glashow-Poisson average energy (thin magenta line) is an upper bound to the more realistic energy presented in this work (thick blue line), which has the expected staircase shape, where in each step the 80\% of the energy is radiated away. In addition, after exactly one mean free path $x=l(E_i)$, the particle has one-half of the initial energy, that is, the average energy loss in each decay times the probability of a decay after a mean free path, as the probability of two or more decays in the first mean free path is negligible  ($\simeq k\, (1-\mathcal{P}(0; l (E_i))) = k\,(1-1/e)\simeq 0.5\,$). On the other hand, the agreement between both models for long enough distances is specious and due to the scaling of the graph; a logarithmic vertical scaling would show (as we have checked) that they do not approximate each other, in accord with what was discussed in Sec.~\ref{sec:CGP}.

\section{Fluctuations in each decay}
\label{sec:each_decay}
\begin{figure}
\centerline{\includegraphics[width=0.47\textwidth]{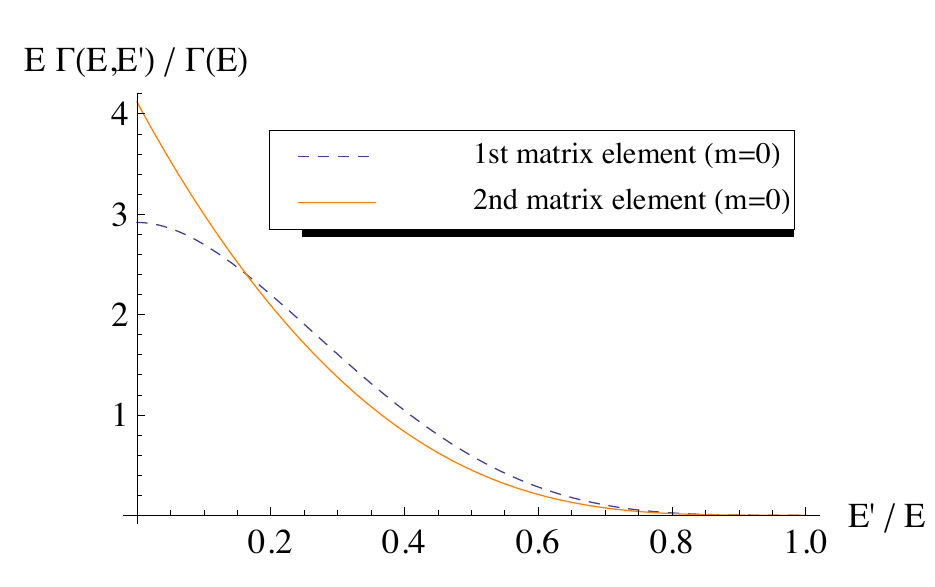}}
\caption{Differential decay rate for two of the dynamical matrix elements reported in Ref.~\cite{Carmona:2012} in the case of energy-independent velocity ($E=p\, (1+\eta_0)$). $E$ and $E'$ are, respectively, the energies of the incoming and outgoing neutrinos.}
\label{fig:Gamma}
\end{figure}

In addition to the fluctuations in the number of disintegrations, there is another source of uncertainty due to the range of energies one particle may lose in each decay, in other words, due to the deviation of the distribution $\Gamma (E, E')/\Gamma (E)$ from a Dirac delta function. For the sake of clarity and succinctness, let us focus on electron-positron pair radiation by superluminal neutrinos in the specific case of an energy-independent speed; see Fig.~\ref{fig:Gamma}. The mean energies for the outgoing neutrinos are $\bar{E}'=0.22\,E$ and $\bar{E}'=0.19\,E$ (equivalently, $k=0.78$ and $k=0.81$, see Eq.~(\ref{energyradiated})) for the first and second dynamical matrix elements of Ref.~\cite{Carmona:2012}, respectively. And the standard deviations are $0.34$ and $0.15$, respectively. As another significant feature of these differential decay widths, the probability that the outgoing neutrino has half or more of the incoming neutrino's energy is just $0.067$ and $0.050\,$.

\vspace{1.1mm}

Nevertheless, even though the probability distributions $\Gamma (E, E')/\Gamma (E)$ for the reaction $\nu_l\rightarrow \nu_l\, e^{+}\, e^{-} $ are not especially concentrated around any specific point, this does not have a strong impact on the final energy uncertainty because this is not very sensitive to variations in $k$ for a given distance of propagation and initial energy. An especially simple way to see it is to compare the terminal energy for two different values of $k$
\begin{equation}
\label{diff_k}
\frac{E_t (k)}{E_t (k')}=\left(\frac{k'}{k}\right)^{1/(5+3m)}\, .
\end{equation}
This, together with the low probability of losing less than the 50\% of the energy in each process for this specific reaction, makes the fluctuations in the final energy due to the uncertainties in the energy loss in each process small. In processes involving other particles, in which the probability of radiating a tiny amount of energy in a decay is not negligible, the fluctuations will be much larger.

\section{Generality of the result: matrix elements and dispersion relations }
\label{sec:generality}

We have so far focused on pair radiation by superluminal neutrinos with an energy-independent speed ($m=0$) to be concise. This is by far the most studied case; see, for example, Refs.~\cite{Cohen:2011,Stecker:2013,Borriello:2013,Cowsik:2012}. Nonetheless, one might wonder how general the preceding results are. We will address this issue in this section.

We have checked that the comparison between the Cohen-Glashow-Poisson model and the one we are presenting in this paper is not qualitatively affected by changes in the dispersion relation (changes in $m$) and the corresponding small modification of $k$ according to Fig.~\ref{fig:k_m}.

\begin{figure}[t]
   \begin{minipage}[c]{8.5cm}\hspace*{-6mm}
     \includegraphics[width=8.8cm,clip=,draft=false]{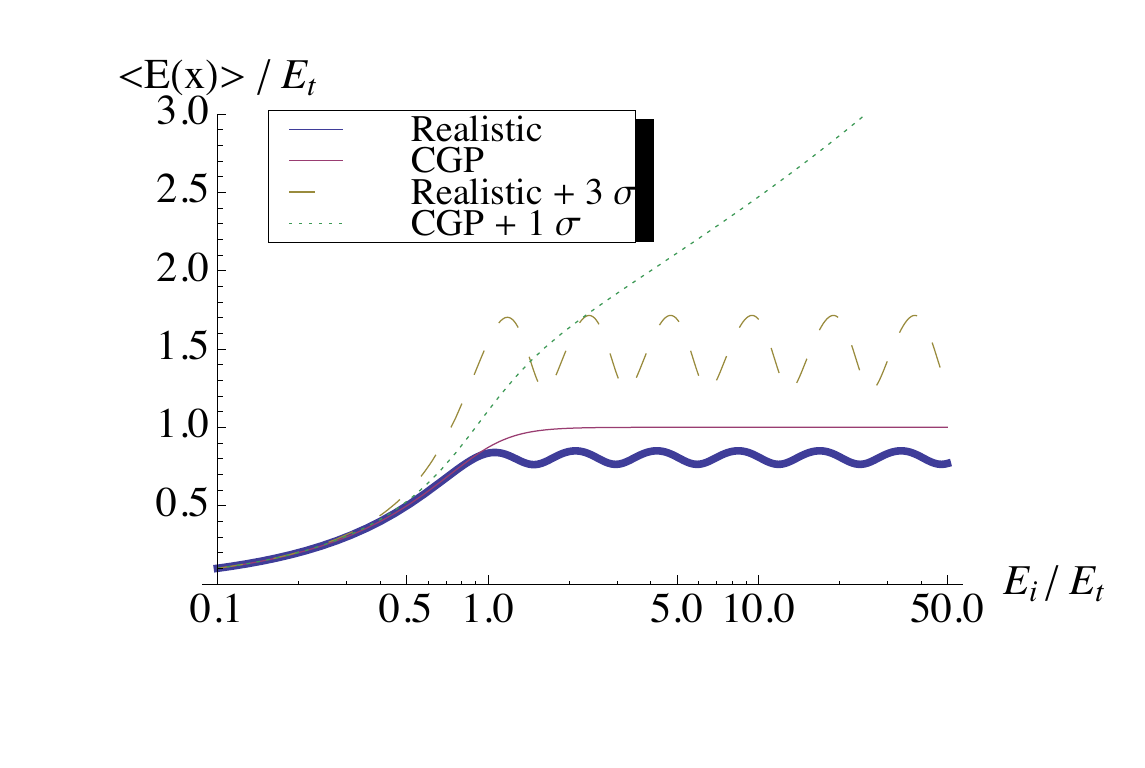}
   \end{minipage}
   \begin{minipage}[c]{8.7cm}\hspace*{-6mm}
     \includegraphics[width=8.8cm,clip=,draft=false]{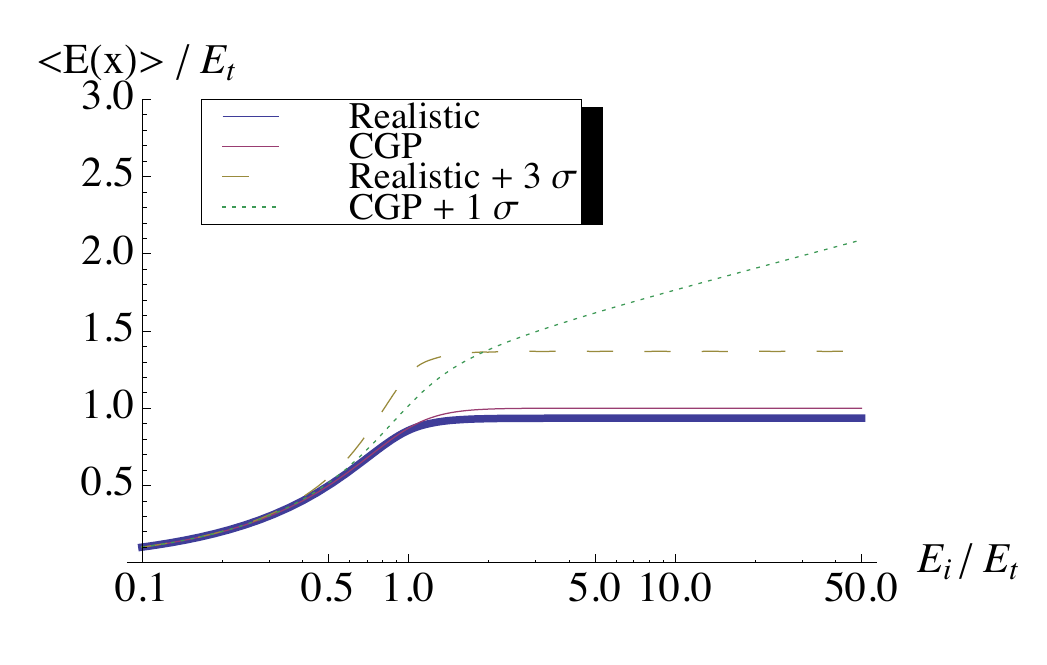}
   \end{minipage}
\caption{Average final energy after a given distance travelled (in terms of the terminal energy for this given distance)  against initial energy (in terms of terminal energy as well) for an energy-independent velocity  ($m=0$) and for $k=0.5$ (above) and $k=0.2$ (below). The thin magenta line represents the Cohen-Glashow-Poisson model, while the thick blue one represents the realistic model presented in this work.}
\label{fig:k=0.5}
\end{figure}

By considering a larger range of values of $k$, we expect to illustratively cover other possible decays, such as some of those mentioned in the introduction of the present work, keeping in mind that there are peculiarities in each of them. For example, photon and neutrino splitting require dispersion relations with energy-dependent speed, photon production by neutrinos is a radiative (loop-suppressed) decay, and superluminarity is needed to create charged lepton pairs.  We, however, remark that $k$ enters through Eq.~(\ref{energyradiated}), and the single condition for this equation to work is that the differential cross section verifies 
\begin{equation}
\Gamma (E, E')\propto E^{\alpha}\, f(E'/E)\, ,
\end{equation}
with $f$ an arbitrary function and $\alpha$ a real number.\footnote{Note that $\alpha$ is nothing more than $4+3\,m$, according to the notation we have used in other sections.} This does not seem a restrictive requirement to fulfil in a decay, provided that energies are high enough  to neglect masses. With this motivation, let's see how the value of $k$ affects the results of this section. In Fig.~\ref{fig:k=0.5} we can see the value of the final energy for $k=0.5$ (above) and $k=0.2$ (below), to be contrasted with Fig.~\ref{fig:Efinal_Einitial}  ($k=0.8$). Some comments are in order: first, and most important, we can check that the Cohen-Glashow-Poisson average energy is an approximation to the realistic one for $k\ll1$ (low energy loss in each reaction, high rate of emission), as we expected from the discussion at the end of Sec.~\ref{sec:CGP}; second, the lower the value of $k$ is, the lower the fluctuations are.

\section{Summary and discussion}

Given the lack of experimental data suggesting energy loss mechanisms due to Lorentz invariance violation, let us contrast the Monte Carlo simulation in Ref.~\cite{Maccione:2011} with the analytical model presented in this paper. The spectrum of that simulation for the energy-independent speed case ($n=2$ in the notation of Ref.~\cite{Maccione:2011}), in which only the process $\nu_l\rightarrow \nu_l\, e^{+}\, e^{-} $ contributes to the energy loss, may be summarized as having a bump between about $7$ and $11$~GeV (with a maximum in 9~GeV) and a tail that extends up to almost 30~GeV and has an amplitude about 10\% of the amplitude of the bump. In this simulation, it is assumed that the parent neutrino's energy is evenly distributed among the daughter neutrino and radiated electron-positron pair in each decay, which means $k=2/3$, and the average initial energy is about 28 GeV. Even though, according to Fig.~\ref{fig:k_m} for $m=0$, this value of $k$ is slightly underestimated for an energy-independent velocity, this should not affect the spectrum very much, as the authors of Ref.~\cite{Maccione:2011} claim and we also argued in Sec.~\ref{sec:each_decay}. The terminal energy for $k=2/3$ and $x=730$~km, the distance used in the simulation, is $E_t=13$~GeV. As we will not use the specific shape of the initial spectrum adopted in the simulation, by no means will we attempt to make a careful comparison, but only an exploratory one. In our model, for $k=2/3$, a particle with initial energy of 28 GeV has a negligible probability  of not decaying ($\sim 10^{-4}$~\%) and a probability of approximately 95\% of decaying only once; therefore, the final energy is very concentrated around $\approx 1/3\cdot 28$~GeV~$\approx 9$~GeV, in perfect agreement with the Monte Carlo simulation (note that it seems that the fluctuations in each decay have been neglected in the simulation, i.e., that neutrinos always lose $2/3$ of their energy in each process). However, neutrinos with initial energies of 18~GeV and 55~GeV have a probability of 20\% of decaying zero times and once, respectively, giving rise to  final energies of 18~GeV, also in agreement with part of the tail of the simulation. However, according to our model, the tail should have a sharper cut-off at about $1.7\cdot E_t\simeq 22$~GeV (the probability that neutrinos with initial energies larger than 22~GeV and 66~GeV decay zero times and once, respectively, is equal to or lower than 1\%), regardless of how large the initial energy is (see Fig.~\ref{fig:Efinal_Einitial}, which is very similar to the one with $k=2/3$). This difference is very significant because the sharp cut-off predicted by our model does allow us to place stringent bounds on Lorentz invariance based on energy loss mechanisms.


In summary, we have shown that the Cohen-Glashow formula for the average energy loss is connected with the assumption that the number of processes along the flight path follows a Poisson statistics. Using this relationship, we have computed the fluctuations in the final energy of the particles, showing that they are so large that they  prevent us from placing stringent constraints in Lorentz invariance violations. This had been largely ignored so far. However, we have argued that Poisson statistics does not take into consideration the sharp  change in the decay rate after each decay. To fix this, we have proposed a probability distribution in Eqs.~(\ref{realistic})~and~(\ref{initial_condition}). The Cohen-Glashow equation for the mean final energy is an upper bound to the final energy averaged with the new distribution. Within this distribution, the fluctuations in the number of decays are much more limited, so that, luckily enough, it is possible to place  bounds on Lorentz invariance violations. More specifically, our model gives rise to a cut-off in the spectrum at a few times the terminal energy, regardless of how large the initial energy of the particles is, at least for the pair production by superluminal neutrinos process, where the probability that a neutrino loses less than half of its energy in each decay is very small. We have also showed that the Cohen-Glashow final energy tends to the average final energy computed within the model presented in this paper when the fraction of radiated energy  (the energy loss over the parent particle's energy) is much lower than one (note, however, that for the process $\nu_l\rightarrow \nu_l\, e^{+}\, e^{-} $ considered in Ref.~\cite{Cohen:2011}, the value of this fraction is about~$0.8$).  


\section{Acknowledgments}
I am grateful to S.~Liberati, L.~Maccione, and especially to  J.~M.~Carmona and C.~Romero-Redondo for a careful reading of the manuscript and comments that significantly improved the presentation of this paper. 


\end{document}